# Reaction of the Desulfoferrodoxin from *Desulfoarculus baarsii* with Superoxide Anion. Evidence for a Superoxide Reductase Activity

Murielle Lombard ‡, Marc Fontecave ‡, Danièle Touati §,

and Vincent Nivière ‡*

Running title : Superoxide Reductase Activity of Desulfoferrodoxin.

‡ Laboratoire de Chimie et Biochimie des Centres Redox Biologiques, DBMS-CEA/CNRS/Université Joseph Fourier, 17 Avenue des Martyrs, 38054 Grenoble, Cedex 9, France

§ Institut Jacques Monod, CNRS/Universités Paris 6 et Paris 7, 2 place Jussieu, 75251 Paris, Cedex 05, France

* To whom correspondence should be addressed.
Telephone : 33-(0)4-76-88-91-09.  Fax : 33-(0)4-76-88-91-24.

E-mail : niviere@cbcrb.ceng.cea.fr

SUMMARY

Desulfoferrodoxin is a small protein found in sulfate-reducing bacteria which contains two independent mononuclear iron centers, one ferric and one ferrous. Expression of desulfoferrodoxin from *Desulfoarculus baarsii* has been reported to functionally complement a superoxide dismutase deficient *Escherichia coli* strain. To elucidate by which mechanism desulfoferrodoxin could substitute for superoxide dismutase in *Escherichia coli*, we have purified the recombinant protein and studied its reactivity towards $O_2^-$. Desulfoferrodoxin exhibited only a weak superoxide dismutase activity (20 U $mg^{-1}$) that could hardly account for its antioxidant properties. UV-visible and EPR spectroscopy studies revealed that the ferrous center of desulfoferrodoxin could specifically and efficiently reduce $O_2^-$, with a rate constant of 6-7 x $10^8$ $M^{-1}$ $s^{-1}$. In addition, we showed that membrane and cytoplasmic *Escherichia coli* protein extracts, using NAD(P)H as electron donors, could reduce the $O_2^-$ oxidized form of desulfoferrodoxin. Taken together, these results strongly suggest that desulfoferrodoxin behaves as a superoxide reductase enzyme, and thus provide new insights into the biological mechanisms designed for protection from oxidative stresses.

INTRODUCTION

Desulfoferrodoxin (Dfx) [1] is a small non-sulfur iron protein that has been isolated from several strains of anaerobic sulfate-reducing bacteria (1, 2). Although no enzymatic activity could be associated to Dfx, the physico-chemical properties of its iron centers have been well documented (1, 2, 3). Recently the three dimensional structure of Dfx from *Desulfovibrio desulfuricans* has been solved at 1.9 Å resolution (4). Dfx is a homodimer with a molecular mass of 2 x 14 kDa. The monomer is organized in two protein domains, each with a specific mononuclear iron center named center I or center II. Center I contains a mononuclear ferric iron coordinated by four cysteines in a distorted rubredoxin-type center. Center II has a ferrous iron with square pyramidal coordination to four nitrogens from histidines as equatorial ligands and one sulfur from a cysteine as the axial ligand (4). The midpoint redox potentials have been reported to be 2 to 4 mV for center I and 90 to 240 mV for center II (2, 3). The high redox potential value for center II explains the stability of the ferrous ion in the presence of oxygen.

Initially, the structural *dfx* gene was cloned and sequenced from *Desulfovibrio vulgaris* Hildenborough and was named *rbo* (5). *rbo* was found upstream of the rubredoxin gene, forming an operon. The encoded 14 kDa protein was tentatively named rubredoxin oxidoreductase (Rbo) because it was likely to function in oxidation-reduction with rubredoxin as a redox partner (5).

Independently, a protein isolated from *D. desulfuricans* and *D.vulgaris* and named Dfx, was found to be encoded by the *rbo* gene (1, 2, 6). However, up no now, Dfx did not show any evidence for a rubredoxin oxidoreductase activity and its physiological role remains unclear. Consequently, the name of the corresponding gene changed from *rbo* to *dfx*.

Recently, Pianzzola et al. attempted to clone the *sod* gene from the sulfate-reducing bacteria *Desulfoarculus baarsii* by functional complementation of a SOD-deficient mutant of *Escherichia coli* (7). They actually found a complementing gene, showing high sequence identity with *dfx* from *D. vulgaris*. However, although expression of *dfx* could fully complement the SOD phenotype, no SOD activity could be detected *in vivo*, raising the question of the functional basis for the successful complementation (7, 8). Further, it has been shown that deletion of the *dfx* gene increased the oxygen sensitivity of *D. vulgaris* when exposed to transitory aerobic conditions (9). This strongly suggested that, in these anaerobic bacteria, Dfx plays a role in the defense against oxidative stress.

In the present work, we have purified the recombinant Dfx from *D. baarsii* and we have investigated the mechanism by which Dfx could trap superoxide and thus replace superoxide dismutase in *E.coli*.

EXPERIMENTAL PROCEDURES

*Biochemical and chemical reagents*. Xanthine, hypoxanthine, potassium ferricyanide NADH, NADPH, nitro blue tetrazolium (NBT) and potassium superoxide were from Sigma. Anhydrous $Me_2SO$ was from Acros Organics. $H_2O_2$, 30% solution in water, was from Aldrich. Xanthine oxidase Grade IV from milk (0.24 U/mg), catalase from *Aspergillus niger* (6600 U/mg), cytochrome c from bovine heart, CuZn-SOD from bovine erythrocytes (5800 U/mg) and Fe-SOD from *Escherichia coli* (5500 U/mg) were from Sigma.

*Strain and plasmid.* E.coli strain QC774 ($\Phi$(*sodA::lacZ* )49 $\Phi$(*sodB::kan* ) 1-$\Delta$2, $Cm^r$ $Km^r$) transformed with plasmid pMJ25 (7) was used for over-expression of the *dfx* gene. pMJ25 contains the *D.baarsii dfx* gene under the control of the IPTG inducible *tac* promoter.

*Overproduction and purification of Dfx*. E.coli QC774 pMJ25 cells were grown aerobically at 37 °C in M9 minimal medium complemented with 0.4% glucose, 2 µg/ml thiamin, 1 mg/ml casamino acid, 2 mM IPTG, 1 mM $FeSO_4$ 7 $H_2O$ and 200 µg/ml ampicilin (12 x 1000 ml in 2 l Erlenmeyer flask). Growth was performed overnight until the culture reaches an OD 600 nm of about 2.3. All following operations were carried out at 4 °C and pH 7.6. The cells were collected by centrifugation. The cell pellet (50 g, wet weight) was suspended in 150 ml of 0.1 M Tris/HCl and sonicated. After ultracentrifugation at 45,000 rpm during 90 min in a Beckman 50.2 Ti rotor, the supernatant was used as soluble extract for further purification. Soluble extract (160 ml, 800 mg of protein) was treated with streptomycin sulfate (final concentration 2% w/v) during one hour of stirring and then centrifuged for 20 min at 14,000 g. The supernatant was treated with 2-3 mg of pancreatic Dnase for 15 min and then precipitated with ammonium sulfate (final concentration 80 % w/v). After centrifugation, 20 min at 14,000 g, the pellet was dissolved in 0.1 M Tris/HCl (10 ml, 80 mg/ml) and the solution was loaded onto an ACA 54 column (360 ml) equilibrated with 20 mM Tris/HCl. Proteins were eluted at a flow rate of 0.4 ml.min$^{-1}$ with the same buffer. A pink fraction corresponding to the volume of elution of low molecular weight protein was collected in a total volume of 57 ml (2 mg/ml). At this stage, the protein solution had already a distinct visible spectrum which resemble to that of Dfx. The ratio $A_{280nm}/A_{503nm}$ was 7. SDS-PAGE on this fraction exhibited three major protein bands located at about 16, 14 and 10 kDa. Protein fractions of 10 mg were then further chromatographied using a Bio-Rad Biologic system on an anion exchange column, Uno Q (Bio-Rad), equilibrated with 10 mM Tris/HCl. A linear gradient was applied (0-0.2 M NaCl) in 10 mM Tris/HCl, with a flow rate of 1 ml.min$^{-1}$ during 30 min.

*Analytical determination*. SDS-PAGE polyacrylamide gels (15% polyacrylamide) were done according to Laemmli (10). The gels were calibrated with the Pharmacia low molecular weight markers. The native molecular mass of the protein was determined with a Superdex 75 gel filtration column (120 ml, Pharmacia) equilibrated with 25 mM Tris/HCl, pH 7.6 and 150 mM NaCl using a flow rate of 0.4 ml.min$^{-1}$. Bovine serum albumin (66 kDa), ovalbumin (45 kDa), trypsine inhibitor (20.1 kDa) and cytochrome c (12.4 kDa) were used as the markers for molecular mass. The void volume was determined with ferritin (450 kDa). Protein concentration was determined using the Bio-Rad protein assay reagent (11) with bovine serum albumin as a standard. Protein-bound iron was determined by atomic absorption spectroscopy. UV-visible spectra were recorded on a Varian Carry 1 spectrophotometer using 1 cm path quartz cuvette. EPR measurements were made on a Bruker EMX 081 spectrometer equipped with an Oxford Instrument continuous flow cryostat.

*N-terminal sequence analysis*. The proteins were separated on SDS-PAGE and then transferred on a ProBlot$^{TM}$ membrane (Applied Biosystem) as described by the manufacturer. NH$_2$-terminal amino acid sequence determination was performed using an Applied Biosystems gas phase separator model 477A with on-line analysis of the phenylthiohydantoin derivatives.

*Mass spectrometry*. Mass spectra were obtained on a Perkin-Elmer Sciex API III+ triple quadrupole mass spectrometer equipped with a nebulizer assisted electrospray source (ionspray) operating at atmospheric pressure.

*Assays for SOD activity*. The SOD activities were evaluated using the cytochrome c reduction assay modified from McCord and Fridovich (12). The assay was performed at 25 °C in a 3 ml of reaction buffer (50 mM potassium phosphate, pH 7.6) containing 22 µM cytochrome c, 200 µM xanthine, 500 U/ml catalase and an amount of xanthine oxidase which gives an initial rate of $\Delta A_{550nm}$

= 0.025 per min in the absence of SOD activity. Reduction of ferricytochrome c was followed at 550 nm, and rates were linear for at least 4 min. One unit of SOD is defined as the amount of protein which inhibits the rate of the reduction of ferricytochrome by 50%. The SOD activities were also evaluated using another test, the NBT reduction technique modified from Beauchamp and Fridovich (13). The assay was performed at 25 °C in a 3 ml of reaction buffer (50 mM Tris/HCl, pH 7.6) containing 22 µM NBT, 200 µM xanthine, 500 U/ml catalase and an amount of xanthine oxidase which gives an initial rate of $\Delta A_{560nm}$ = 0.0165 per min in the absence of SOD activity. Reduction of NBT was followed at 560 nm during 0.5 min, in order to keep absorbance changes fairly low and avoid precipitation of formazan. One unit of SOD is defined as the amount of protein which inhibits the rate of the reduction of NBT by 50%. In our experimental conditions, using the same commercial CuZn-SOD preparation, NBT test gave a 4 fold more elevated SOD specific activity compared to that determined with the cytochrome c reduction test.

*Preparation of cell extracts, reduction of center II*. Aerobic cultures of *E.coli* strain QC774 pMJ25 were grown in Luria-Bertani (LB) medium at 37 °C and were harvested at about 0.3 $OD_{600nm}$. All the following steps were performed at 4 °C. Cultures were centrifuged, washed in cold 50 mM potassium phosphate buffer pH 7.8 and resuspended in about 3 % of the original culture volume in the same buffer. The cells were then sonicated and centrifuged for 20 min at 6,000 rpm, to remove cell debris. The supernatant, which contained both cytosolic and membrane material, was fractionated by a centrifugation at 45,000 rpm (Beckman, 70.1 Ti rotor) for 2 hours. The supernatant cytosolic fraction (2.3 mg/ml), was removed and stored at -80 °C. The pelleted membranes vesicles were resuspended in 50 mM potassium phosphate buffer pH 7.8 plus 100 mM NaCl and recentrifuged. The resulting pellet was resuspended in 0.2 % of the original culture volume in 50 mM potassium phosphate buffer pH 7.8. The membrane fractions were stored at 0 °C and used within a week.

*Oxidation of Dfx*. $KO_2$ stock solutions were prepared as followed. $Me_2SO$ was dried over 3-Å molecular sieves. Potassium superoxide was dissolved in anhydrous $Me_2SO$, under a dry atmosphere of argon. No crown ether was added. The concentration of superoxide was determined by using its absorbance in the UV with an extinction coefficient of 2086 $M^{-1}$ $cm^{-1}$ at 260 nm. $Me_2SO$ stock solutions of potassium superoxide (1-2 mM) were prepared immediately before each experiment. The fully oxidized form of Dfx was also obtained by addition of 2 fold molar excess of potassium ferricyanide to the purified Dfx. Complete oxidation of center II was verified by UV-visible spectrophotometry. Excess of potassium ferricyanide was eliminated by washing with 50 mM Tris/HCl pH 7.6, using a Centricon pM30.

RESULTS

*Purification of the recombinant Dfx*

In order to avoid any possible contamination by superoxide dismutase (SOD) activities from the host strain, the *dfx* gene was overexpressed in the *E.coli* QC 774 strain, in which the *sodA* and *sodB* genes were insertionally inactivated (7). A two-step purification protocol, with a gel filtration on ACA 54 and anion exchange chromatography on Uno Q (Bio-Rad) as described in the experimental section, was set up. Samples were analyzed at all steps by SDS-PAGE and UV-visible spectroscopy, since Dfx exhibits characteristic absorption bands responsible for the pink color of the protein (1). During Uno Q chromatography, a pink fraction, eluted at 50 mM NaCl, was collected (fraction A). It contained only two polypeptide bands at 14 and 16 kDa with about equal intensities, as shown by SDS-PAGE analysis. Another pink fraction was eluted at a slightly higher NaCl concentration (fraction B). In addition to the 14 and 16 kDa polypeptides, fraction B contained also a major polypeptide of 10 kDa, as shown by SDS-PAGE.

The 14 and 16 kDa polypeptides had the same PERLQVYKCE N-terminal sequence, identical to the N-terminal translated sequence of the *D.baarsii dfx* structural gene, without the N-terminal Met residue (7). Furthermore, they had the same mass, as shown by electrospray mass spectrometry analysis, with only ions detected at 14,028±2 Da, confirming the absence of the N-terminal Met residue. These data suggested that the 14 and 16 kDa bands on SDS-PAGE originated from the same Dfx polypeptide. In fact, the proportion of the two polypeptide bands was found to be correlated to the presence of the reducing agent, DTT or ß-mercaptoethanol, during electrophoresis (data not shown). From 800 mg of soluble extracts, 23 mg of pure Dfx (fraction A) were obtained.

The 10 kDa protein, present only in fraction B, exhibited the N-terminal ADAQKAADNKKPVN sequence, which is identical to the N-terminal sequence of the mature form of HdeA (14). HdeA is a highly abundant periplasmic protein of *E.coli* of unknown function (14). Fraction B appeared then not homogeneous and was not further characterized.

Finally, during Uno Q chromatography, a protein peak, eluted from the column at 90 mM NaCl, also contained the 14 and 16 kDa proteins together with other minor contaminants. However, this fraction was colorless and did not display the UV-visible spectrum characteristic of Dfx. It thus probably contained the apo form of Dfx and was thus discarded.

Gel-filtration experiments on Superdex 75 column with the native recombinant Dfx, as described in the Experimental Procedure, gave an apparent molecular mass of 27,000 Da (data not shown) showing that Dfx from *D. baarsii* is a homodimer.

*Iron Content/Absorption Spectra of the recombinant Dfx*

The iron content of Dfx was determined by atomic absorption spectrophotometry. A value of 1.97 Fe/polypeptide chain (14,026 Da) was found, suggesting that both center I and center II were fully metallated.

As shown in Figure 1, the UV-visible spectrum of the as-isolated Dfx exhibits absorptions at 370 and 503 nm contributed by the ferric iron from center I (1, 2). The ratio $A_{280nm}/A_{503nm}$ was 4.5. The value of the molar extinction coefficient at 503 nm was determined to be 4,400 $M^{-1}$ $cm^{-1}$ (center I). This value was similar to the corresponding value reported for *D. desulfuricans* and *D.vulgaris* Dfx (1, 2). Figure 1 also shows the spectrum of the protein treated with an excess of potassium ferricyanide. It is characteristic of the grey form of Dfx, with absorptions contributed by the ferric irons from both centers I and II (3). In the inset, the difference spectrum provides the contribution of the oxidized center II with absorption bands centered at 644 and 330 nm (3). The value of the molar coefficient at 644 nm was found to be 1,900 $M^{-1}$ $cm^{-1}$ (center II).

*Epr spectroscopy analysis*

The EPR spectrum of the protein, recorded at 4°K, displays resonances at g = 7.7, 5.7, 4.1 and 1.8 (data not shown). It is similar to that reported for the pink form of Dfx from *D. desulfuricans* (1) and *D.vulgaris* (2). It is typical for a distorted $FeS_4$ center (S = 5/2), assigned to the center I (1, 2). When the EPR spectrum was recorded at 10 °K, the intensity of the g = 7.7 feature decreased while that of the g = 5.7 feature increased (data not shown). This is consistent with the former being derived from the ground state and the latter from an excited state, as previously reported (1). The 4 °K EPR spectrum of the oxidized protein presents a signal at g = 4.3, in addition to the features at g = 7.7, 5.7, and 1.8 (data not shown). This spectrum is similar to that reported for the ferric form of Dfx from *D. desulfuricans* (3) and from *D.vulgaris* (2), and the signal at g = 4.3 was attributed to oxidized center II (2, 3).

*Dfx is not a superoxide dismutase*

The ability of Dfx to catalyze the dismutation of $O_2^-$ was assayed from its inhibitory effect on the reduction of cytochrome c by superoxide, generated by the xanthine-xanthine oxidase system (12). In Fig. 2A are shown the traces of the reduction of cytochrome c in the presence of different amounts of Dfx. With up to 11 µg of Dfx, almost no inhibition of cytochrome c reduction could be observed (data not shown). However, larger amounts of Dfx resulted in a two-phase kinetics. An initial lag period, corresponding to a complete inhibition of cytochrome c reduction, was now observed (Fig. 2A). Duration of the lag period was found roughly proportional to the amount of Dfx added in the test cuvette (Inset of Fig 2A). In the second phase of the reaction, formation of reduced cytochrome c appeared linear with time, but with a slope that slightly decreased with increased Dfx concentration (Fig. 2A).

When Dfx was pretreated with the superoxide-generating system (xanthine-xanthine oxidase) and then assayed for inhibition of cytochrome c reduction, the lag phase could not be observed anymore (Fig. 2B). However, increased amounts of the preincubated Dfx inhibited the reduction of cytochrome c. The addition of 100 μg of preincubated Dfx resulted in 50% inhibition of cytochrome c reduction (Inset Fig. 2B), a value corresponding to a specific SOD activity for the preincubated Dfx of 20 U mg$^{-1}$. A comparable value of specific SOD activity was found from the linear second phase of the kinetic of Fig. 2A (data not shown).

When a Dfx solution (39 μM in 50 mM Tris/HCl, pH 7.6) was pretreated with a 5 molar excess of $O_2^-$ ($KO_2$ dissolved in $Me_2SO$), in the presence of 500 U/ml catalase, comparable results to Fig. 2B were obtained (data not shown).

Similar results were obtained using another assay, the so-called NBT reduction assay (13). The superoxide-dependent reduction of NBT by the xanthine-xanthine oxidase system was inhibited by large amounts of Dfx (data not shown). Kinetics of reduction were linear for at least 0.5 min and no lag time was observed in the presence of Dfx. 9 μg of Dfx induced a 50% inhibition of reduction of NBT (data not shown), a value corresponding to a specific SOD activity of 25 U mg$^{-1}$ in the cytochrome c assay. Preincubation of a concentrated Dfx solution with the xanthine-xanthine oxidase system (as reported in Fig. 2B) before the assay, gave a comparable value of the specific SOD activity (data not shown).

*Superoxide oxidizes efficiently Dfx center II*

The results from the cytochrome c and NBT assays suggested that Dfx did not exhibit a significant superoxide dismutase activity. However, the observation, in the cytochrome c assay, of a lag period proportional to the amount of added Dfx (Fig. 2A) suggested that, during this period, $O_2^-$ reacted with Dfx rather than with cytochrome c. As shown in Fig. 3A, incubation of Dfx with the $O_2^-$ generating system (xanthine-xanthine oxidase plus catalase) induced an increase of the protein absorbance in the 600-700 nm range. Difference spectra clearly showed a specific oxidation of center II, with the appearance of the band centered at 644 nm (Inset Fig. 3A). Under these conditions, after 10 minutes incubation, the oxidation of center II was complete. Longer incubation time did not further modify the spectrum of the fully oxidized Dfx (data not shown). The same results were obtained in the absence of catalase (data not shown).

Fig. 3B shows the effect of successive additions of stochiometric amounts of $O_2^-$ ($KO_2$ dissolved in $Me_2SO$) on the visible spectrum of Dfx, in the presence of catalase. Spectra changes occurred during the mixing time. A 4-fold molar excess of $O_2^-$ induced a complete and selective oxidation of center II, as shown by the increase of the band at 644 nm (Inset Fig. 3B). Addition of an equivalent amount of $Me_2SO$ without $KO_2$ had no effect on the visible spectrum of Dfx (data not shown). Comparable results were obtained in the absence of catalase (data not shown). Addition of a molar excess of sodium ascorbate to the $O_2^-$ fully oxidized Dfx restored the original visible spectrum of Dfx (data not shown).

The fact that center II was oxidized by $O_2^-$ has been further confirmed by EPR spectroscopy. The EPR spectrum recorded at 4 °K of a 300 μM Dfx solution in 50 mM Tris/HCl, pH 7.6 treated with a 5-fold excess of $O_2^-$, as previously described, was similar to that of Dfx oxidized with potassium ferricyanide, with g values at 7.7, 5.7, 1.8 and 4.3 (data not shown).

*Kinetic parameters associated with oxidation of center II by $O_2^-$*

In order to estimate the rate constant for the oxidation of center II by $O_2^-$, we have used a methodology which has been developed in the case of several dehydratases, such as aconitase and fumarase A, which are known to react with $O_2^-$ very rapidly (15, 16, 17).

The kinetics of the oxidation of center II by $O_2^-$ was followed spectrophotometricaly at 644 nm, in the absence or in the presence of different amounts of CuZn- or Fe-SOD. $O_2^-$ was generated by the xanthine-xanthine oxidase system. As shown in Fig. 4A and 4B, in the absence of SOD, oxidation of center II by $O_2^-$ was linear with time during the first 2 min and was complete after about 5 min reaction. From these data, the initial rate of oxidation of center II was calculated to be 4.2 nmol/min. Considering that under these conditions the production of $O_2^-$ by the xanthine-xanthine oxidase system can be estimated at 4.3 nmol $O_2^-$ /min (measured as the SOD inhibitable reduction of cytochrome c, as reported in (17)), it was suggested that under these conditions, almost all the enzymatically produced $O_2^-$ was used to oxidize center II. The spontaneous dismutation of $O_2^-$ would be therefore relatively insignificant, in agreement with an efficient trapping of $O_2^-$ by Dfx.

In the presence of high amounts of CuZn-SOD (Fig. 4A) or Fe-SOD (Fig. 4B) the rate of oxidation of center II was strongly decreased. Under these conditions, the rate of reaction of $O_2^-$ with Dfx can be described by Eq. 1 :

$$v_{ox} = k_{Dfx} [\text{Dfx}][O_2^-] \qquad (\text{Eq. 1})$$

where $v_{ox}$ and $k_{Dfx}$ are the rate and the second order rate constant of oxidation of Dfx by $O_2^-$, respectively. From the data of Fig. 4, the value for $k_{Dfx}$ can be determined as follows. During the early phase of the reaction :

$$\frac{d[O_2^-{}_{(XO)}]}{dt} - \frac{d[O_2^-{}_{(Dfx)}]}{dt} - \frac{d[O_2^-{}_{(SOD)}]}{dt} = 0 \qquad (\text{Eq. 2})$$

where $d[O_2^-{}_{(XO)}]/dt$ is the rate of synthesis of $O_2^-$ by the xanthine-xanthine oxidase system, $d[O_2^-{}_{(Dfx)}]/dt$, the rate of disappearance of $O_2^-$ due to the reaction with Dfx and $d[O_2^-{}_{(SOD)}]/dt$, the rate of disappearance of $O_2^-$ due to the reaction with SOD. No term of spontaneous decomposition of $O_2^-$ is included in Eq. 2 since it would be negligible, as mentioned above. Rate equations for the production of $O_2^-$ by the xanthine-xanthine oxidase system (XO) and reaction of $O_2^-$ with Dfx and SOD are :

$$\frac{d\,[O_2^-{}_{(XO)}]}{dt} = k_{xo}\,[XO] \qquad (Eq.\,3)$$

$$\frac{d\,[O_2^-{}_{(Dfx)}]}{dt} = k_{Dfx}\,[Dfx][O_2^-] \qquad (Eq.\,4)$$

$$\frac{d\,[O_2^-{}_{(SOD)}]}{dt} = k_{SOD}[SOD][O_2^-] \qquad (Eq.5)$$

Combining Eq. 2 and the expressions of Eq. 3, 4 and 5 would give the following expression for the steady-state concentration of $O_2^-$ in the presence of Dfx and SOD :

$$[O_2^-] = \frac{k_{xo}\,[XO]}{k_{Dfx}\,[Dfx] + k_{SOD}\,[SOD]} \qquad (Eq.\,6)$$

Eq. 1 and 6 could be combined to give Eq. 7 :

$$v_{ox} = k_{Dfx}\,[Dfx]\,\frac{k_{xo}\,[XO]}{k_{Dfx}\,[Dfx] + k_{SOD}\,[SOD]}$$

$$\frac{1}{v_{ox}} = \frac{1}{k_{xo}\,[XO]} + \frac{k_{SOD}}{k_{xo}\,[XO]\,k_{Dfx}\,[Dfx]}\,[SOD] \qquad (Eq.7)$$

Insets of Figures 4A and 4B showed a linear plot of the reciprocal of the initial rate of oxidation of center II ($v_{ox}$) versus CuZn- and Fe-SOD concentration respectively, according to Eq. 7. Under these conditions, when the initial rate of the oxidation of center II is decreased by 50% due to the competition of SOD for $O_2^-$, Eq. 4 and 5 can be rearranged to give :

$$k_{SOD}\,[SOD] = k_{Dfx}\,[Dfx] \qquad (Eq.8)$$

The concentration of CuZn- and Fe-SOD which decrease by 50% the rate of oxidation of center II were then graphically determined from the insets of Fig. 4A and 4B. Values of 6.5 and 38.5 µM of CuZn- and Fe-SOD respectively were found. Taking into account the known second order rate constant of the reaction of $O_2^-$ with CuZn-SOD and with Fe-SOD at low [$O_2^-$], $2\ 10^9$ $M^{-1}$ $s^{-1}$ (18, 19) and $3.25\ 10^8$ $M^{-1}$ $s^{-1}$ (19) respectively, the second order rate constant of the oxidation of center II by $O_2^-$ can be now calculated using Eq. 8. Values of $6.8\ 10^8$ $M^{-1}$ $s^{-1}$ and $6.5\ 10^8$ $M^{-1}$ $s^{-1}$ in the experiments using CuZn- and Fe-SOD respectively were obtained.

*Center II of Dfx is slowly oxidized by $H_2O_2$*

The experiments presented above have been set up in the presence of catalase in order to eliminate a possible effect of $H_2O_2$, the superoxide reduction and dismutation product. The ability of $H_2O_2$ to react with Dfx and to oxidize center II was nevertheless tested spectrophotometrically. When 100 µM of Dfx, in 50 mM Tris/HCl, pH 7.6, was incubated with 1 mM $H_2O_2$, the UV-visible spectrum exhibited an increase of the absorbance in the 600-700 nm range, during the first 5 min reaction (data not shown). Difference spectra clearly showed a complete oxidation of center II, with appearance of the band centered at 644 nm (data not shown). In a presence of 500 U/ml of catalase, no modification of the UV-visible spectrum could be observed (data not shown). The kinetic of the oxidation of center II (100 µM Dfx, in 50 mM Tris/HCl, pH 7.6) by 0.8, 1, 1.5 or 2 mM $H_2O_2$ was followed spectrophotometrically at 644 nm, at 25 °C. In all cases, oxidation of center II was found to follow a pseudo first order kinetic (data not shown). A value of the second order rate constant of oxidation of center II by $H_2O_2$ of 45 $M^{-1}$ $s^{-1}$ was determined. This is almost negligible compared to the value of the rate constant of oxidation of center II by $O_2^-$ (see above).

*Reduction of center II by cell extracts*

The results presented above have shown that the reduced form of center II of *D. baarsii* Dfx can transfer one electron to $O_2^-$ very efficiently. In order to provide evidence that such a reaction could be catalytic within the cell, we have examined the capability of *E.coli* cell extracts to reduce the oxidized form of center II, which then could be involved in a new reaction cycle with $O_2^-$.

The fully oxidized form of Dfx was incubated anaerobically with catalytic amounts of cytosolic or membrane cell extracts, in the presence of NADH or NADPH as electron donors (Table I). Time-dependent reduction of Dfx was followed spectrophotometrically, using a diode-array spectrophotometer (data not shown). Both cytosolic and membrane cell fractions were found to catalyze electron transfer to the center II of Dfx. Complete reduction of center II was observed in the presence of NADH and membrane fractions or in the presence of NADPH and cytosolic fractions (data not shown). In all cases, no evidence for reduction of center I was observed during the time course of the reduction of center II (data not shown). However, longer incubation of Dfx with NADH and membrane fractions or NADPH and cytosolic fractions led to complete reduction of center I, giving the fully reduced form of Dfx (data not shown).

As illustrated in Table I, the rate of reduction of center II depended both on the electron donor and the cell fraction. NADH and the membrane fraction or NADPH and the cytosolic fraction gave the higher rate of reduction of center II, with specific activities values of 90 and 120 nmol of center II reduced /min/ mg of extract, respectively. On the other hand, NADH and the cytosolic fraction were poorly active with a specific activity value of 20 nmol/min/mg. NADPH and the membrane fraction were also found to reduce center II but with a low specific activity of 20 nmol/min/mg. Taking into account that no NADPH dependent reductase should be associated with the membrane fraction in *E.coli*, this residual NADPH dependent reductase activity could originate from cytosolic contaminant proteins.

DISCUSSION

In order to identify the mechanism by which Dfx from *D.baarsii* could protect a SOD-deficient *E.coli* strain from a superoxide stress, we have investigated the reactivity of Dfx with regard to the superoxide anion.

Our results show that Dfx exhibits a very weak SOD activity (20 U/mg), representing about 0.3 % of the specific activity of a CuZn- or Fe-SOD, assayed under comparable conditions. During this work, a comparable low SOD activity of Dfx from *D. desulfuricans* has also been reported (20). The finding that Dfx did not efficiently catalyze the dismutation of $O_2^-$ is not surprising, since no sequence similarity was found between Dfx and any class of SOD characterized so far (7). Furthermore, no SOD activity could be detected in extracts of SOD deficient *E.coli* strain overproducing Dfx (7, 8).

Whether this low SOD activity could nevertheless account for the functional complementation of the SOD deficient *E.coli* strain when Dfx is expressed within the cell is questionable. Recent results from Gort and Imlay showed that *E.coli* can tolerate only small decreases in SOD content (21) and *E.coli* constitutively synthesizes just enough SOD (Fe-SOD) for protection from endogenous $O_2^-$. One can estimate that, under the overexpression conditions used during the complementation experiments (7), Dfx represents no more than 5% of the total soluble proteins, corresponding to less than 1 U mg$^{-1}$ of SOD activity. Such an amount of SOD activity is certainly too low to protect the cell from endogenous $O_2^-$ (21), thus suggesting another mechanism for the antioxidant properties of Dfx.

Indeed, our results suggest that Dfx could eliminate $O_2^-$ by a different reaction and indicate that the reduced form of center II of Dfx has the potential to reduce $O_2^-$ very efficiently. First, the reduced form of Dfx center II was completely oxidized to the ferric form in the presence of $O_2^-$, as shown by UV-visible and EPR spectroscopy experiments. Second, the reaction was a very fast process: the cytochrome c reduction test showed that $O_2^-$ reacted with reduced Dfx much faster than with cytochrome c (cytochrome $c_{ox}$ + $O_2^-$ → cytochrome $c_{red}$ + $O_2$ , $k$ = 2.6 $10^5$ $M^{-1}$ $s^{-1}$(22)); only a 4-fold molar excess of $KO_2$ was required for complete oxidation of center II (taking into account that the spontaneous dismutation of $O_2^-$ is a very fast process, $HO_2^.$ + $O_2^-$ → $H_2O_2$ + $O_2$ $k$ = 8 $10^7$ $M^{-1}$ $s^{-1}$ at neutral pH (23), reaction of Dfx with $O_2^-$ has to be as fast or faster); large amounts of CuZn- or Fe-SOD were necessary to inhibit the oxidation of Dfx center II by $O_2^-$, allowing to calculate a value for the rate constant of the oxidation of center II by $O_2^-$ of 6-7 $10^8$ $M^{-1}$ $s^{-1}$. This value is comparable to the value of the rate constant determined for SOD ( 2 $10^9$ $M^{-1}$ $s^{-1}$ for CuZn-SOD, 3.25 $10^8$ $M^{-1}$ $s^{-1}$ for Fe-SOD). Third, oxidation of center II appeared to be specific for $O_2^-$. $O_2$ does not oxidize the reduced form of center II, and the rate of oxidation by $H_2O_2$ was found to be negligible compared to the rate of oxidation by $O_2^-$.

The large rate of reduction of $O_2^-$ by center II and the specificity for $O_2^-$ strongly support the idea that this reaction occurs within the cell and could be physiologically relevant.

Whether oxidation of center II by $O_2^-$ could be catalytic within the cell depends on the presence of a cellular system able to reduce the oxidized center II for a new reduction cycle with $O_2^-$. We have found that both the cytosolic and membrane fractions of *E.coli* contained NAD(P)H reductase activities which may fulfill this function. As expected, the membrane reductase(s) were found to be NADH-dependent and the cytosolic reductase(s) were found to be rather NADPH-specific. The values of the specific activities of reduction of center II (Table I) are in the range of specific activities reported in crude extracts for many enzymatic systems in *E.coli*, in agreement with a possible *in vivo* catalytic reduction of center II.

Altogether, these data strongly support a superoxide reductase activity for *D.baarsii* Dfx, as previously hypothesized by Liochev and Fridovich (8), which could account for the functional complementation of the SOD-deficient mutant of *E.coli* strain. Efforts to purify the putative *E.coli* NAD(P)H-dependent Dfx reductases are currently under way and would allow to determine the global kinetic parameters for the reduction of superoxide catalyzed by Dfx.

On the other hand, no obvious function could be assigned to Dfx center I yet. The hypothesis that center I could act as an electron relay between cellular reductases and center II was attractive but is not clearly supported by our results.

*E.coli* extracts did not seem to reduce efficiently Dfx center I in the presence of NAD(P)H as electron donors. In addition, the three dimensional structure of *D.desulfuricans* Dfx indicates a distance of 20 Å between center I and II (4), which hardly supports possible electronic interactions between the 2 redox centers. Further investigations are needed to understand the function of center I in Dfx.

Recently, it has been shown that deletion of *dfx* gene increases the oxygen sensitivity of *D.vulgaris* when exposed transitory to microaerophilic conditions (9). Since *dfx* deletion does not affect growth of *D.vulgaris* under anaerobic conditions, it was proposed that the main physiological function of Dfx is that of an antioxidant protein in *Desulfovibrio* spp. (9). This is in line with the functional complementation by Dfx of *sodA sodB* mutant in *E.coli* (7) and we propose that Dfx could protect cell against oxidative stress by the same mechanism in both *E.coli* and in sulfate-reducing bacteria. What could be then the advantage for a cell to have a mechanism of elimination of superoxide by reduction rather than by catalyzing its dismutation with a SOD enzyme ? That Dfx is the survivor of an ancestral system of $O_2^-$ elimination could be considered, but we would favor a more specific function of Dfx, taking into account the particular redox status in sulfate-reducing bacteria.

Anaerobic bacteria, and in particular sulfate-reducing bacteria, are usually known to be highly sensitive to exposure to air, during which a whole array of enzymes and proteins can be totally inactivated (24). Furthermore, sulfate-reducing bacteria are fully crowded with strongly auto-oxidizable redox proteins, such as redox carriers (ferredoxin, cytochromes, rubredoxin, desulforedoxin, flavodoxin) or enzymes, like hydrogenases (24). Upon exposure to $O_2$, these proteins are prone to release their electrons, thus inducing a strong superoxide stress (25). Such a process is probably less important in aerobic cells, which have evolved by integrating the electron transport proteins into the membrane in order to minimize such auto-oxidation reactions (22). SOD and catalase have been found in a few sulfate-reducing bacteria (25, 26) and could well account for the good aerotolerance which has been reported in these species (26, 27). However, the presence of Dfx in these bacteria may provide an additional advantage. It is tempting to suggest that Dfx by shuttling the electrons from the auto-oxidizable redox proteins to superoxide preferentially, in a single reaction, could eliminate both superoxide and the source of its production. Another advantage would be that oxidation of redox carriers by Dfx stops as soon as the superoxide stress is over, restoring anaerobic function, without further loss of reducing equivalents. Finally, such a reaction allows these anaerobic bacteria to shut off transitory superoxide production from those redox carriers with no need for sophisticated regulatory systems, such as those found in facultative anaerobes.

ACKNOWLEDGMENTS

We are grateful to Dr. Yves Pétillot for mass spectrometry experiments, to Dr. Jérome Garin and Mathilde Louwagie for N-terminal amino-acid sequence determination and to Dr. Véronique Ducros for atomic absorption spectroscopy. We acknowledge Dr. Stéphane Ménage for helping in EPR experiments and Chantal Falco for technical assistance.

FOOTNOTES

[1] The abbreviations used are: Dfx, desulfoferrodoxin; Rbo, rubredoxin oxidoreductase; SOD, superoxide dismutase; NBT, nitro blue tetrazolium; XO, xanthine oxidase.

FIGURE LEGENDS

Figure 1. Absorption spectra of the recombinant *D.baarsii* Dfx (82.1 µM in 50 mM Tris/HCl pH 7.6). Spectrum of the as-isolated Dfx (—), or treated with 100 µM potassium ferricyanide (- - -). The inset shows difference spectrum of (- - -) minus (—).

Figure 2. Cytochrome c reduction assay in the presence of *D.baarsii* Dfx. Reduction of ferricytochrome c (22 µM) was followed at 550 nm at 25 °C, in the presence of 50 mM potassium phosphate pH 7.6, 200 µM xanthine, 500 U/ml catalase. Reaction was initiated by addition of 0.023 U xanthine oxidase. A. Assay in the presence of various amount of as-isolated Dfx. (○) 0 µg; (□) 22 µg; (◇) 33 µg; (▲) 44 µg and (▽) 66 µg. Arrows indicate the end of the lag time. The inset shows the duration of the lag time as a function of the amount of Dfx added in the assay. B Assay in the presence of various amount of Dfx preincubated with the xanthine-xanthine oxidase system. Preincubation conditions: 39 µM Dfx, 50 mM Tris/HCl pH 7.6, 200 µM hypoxanthine, 0.023 U xanthine oxidase, 500 U/ml catalase, 10 min at 25 °C. (○) 0 µg; (□) 22 µg; (◇) 44 µg and (▲) 66 µg. The inset shows the initial rate of reduction of cytochrome c as a function of the amount of Dfx in the assay.

Figure 3. Effect of $O_2^-$ on the visible spectra of *D. baarsii* Dfx, at 25 °C. A. Xanthine oxidase as an $O_2^-$ generating system. The micro-cuvette (120 µl final volume) contains, 73 µM Dfx, 50 mM Tris/HCl pH 7.6, 400 µM hypoxanthine, 500 U/ml catalase. The reaction was initiated by adding 0.006 U xanthine oxidase. Spectra were recorded at different time interval, from the bottom to the top, at 0, 80, 160, 240, 320, 400 and 480 sec. The inset shows the difference spectra, after each incubation time minus time zero. B. $KO_2$ prepared in $Me_2SO$, as a source of $O_2^-$. The micro-cuvette (200 µl final volume) contains 82 µM Dfx in 50 mM Tris/HCl pH 7.6, 500 U/ml catalase. Successive additions of 82 µM $KO_2$, from a 1.5 mM $KO_2$ stock solution dissolved in 100% $Me_2SO$ (14 M), were performed. After each addition, a spectrum was recorded. From the bottom to the top, no addition, 1 equivalent, 2 equivalents, 3 equivalents, 4 equivalents. The inset shows the difference spectra, after each addition minus no addition.

Figure 4. Kinetics of oxidation of center II from the *D. baarsii* Dfx by $O_2^-$. Oxidation of center II was followed spectroscopically, at 25 °C, by the increase of absorbance at 644 nm. The cuvette contains (1 ml final volume) 19 µM Dfx, 50 mM Tris/HCl pH 7.6, 400 µM xanthine, 500 U/ml catalase and different amounts of CuZn-SOD or Fe-SOD. The oxidation was initiated by adding 0.013 U of xanthine oxidase. A. In the presence of CuZn-SOD. For the shake of clarity, only the following traces are presented: (○) 0 µM; (□) 1.1 µM; (◊) 5.7 µM and (▲) 8.6 µM of CuZn-SOD. The inset shows the reciprocal of the initial velocity of the oxidation of center II as a function of [CuZn-SOD]. B. In the presence of Fe-SOD. For the shake of clarity, only the following traces are presented: (○) 0 µM; (□) 8.6 µM; (◇) 26.0 µM and (△) 34.8 µM of Fe-SOD. The inset shows the reciprocal of the initial velocity of the oxidation of center II as a function of [Fe-SOD].